\documentclass[sn-mathphys-num]{sn-jnl}

\usepackage{graphicx}%
\usepackage{multirow}%
\usepackage{amsmath,amssymb,amsfonts}%
\usepackage{amsthm}%
\usepackage{mathrsfs}%
\usepackage[title]{appendix}%
\usepackage{xcolor}%
\usepackage{textcomp}%
\usepackage{manyfoot}%
\usepackage{booktabs}%
\usepackage{algorithm}%
\usepackage{algorithmicx}%
\usepackage{algpseudocode}%
\usepackage{listings}%

\theoremstyle{thmstyleone}%
\theoremstyle{thmstyletwo}%

\theoremstyle{thmstylethree}%

\begin{document}

\title[Article Title]{Spatio-Temporal Differences in Bike Sharing Usage: A Tale of Six Cities}


\author*[1,2]{\fnm{Shu-ichi} \sur{Kinoshita}}\email{s\_kino@musashino-u.ac.jp}

\author[1]{\fnm{Yuya} \sur{Bando}}\email{nkoa99072@gmail.com}

\author[2,3]{\fnm{Hiroki} \sur{Sayama}}\email{sayama@binghamton.edu}

\affil*[1]{\orgdiv{Department of Mathematical Engineering}, \orgname{Musashino University}, \orgaddress{\street{3-3-3 Ariake}, \city{Koto-ku}, \postcode{135-8181}, \state{Tokyo}, \country{Japan}}}

\affil[2]{\orgdiv{Binghamton Center of Complex Systems}, \orgname{Binghamton University}, \orgaddress{\street{4400 Vestal Parkway East}, \city{Vestal}, \postcode{13902}, \state{New York}, \country{United States}}}

\affil[3]{\orgdiv{Waseda Innovation Lab}, \orgname{Waseda University}, \orgaddress{\street{1-104 Totsukamachi}, \city{Shinjuku-ku}, \postcode{169-8050}, \state{Tokyo}, \country{Japan}}}


\abstract{

This study investigates the spatio-temporal patterns of Bike Sharing System (BSS) usage in six major cities: New York, London, Tokyo, Boston, Chicago and Washington D.C. By analyzing data over a 30-day period with comparable climate and average temperatures, we explored differences in BSS usage between weekdays and weekends in those cities using Jensen-Shannon divergence (JSD) and rank distribution analysis.
Our findings reveal significant temporal differences in BSS usage that were commonly observed in all cities, with weekday patterns dominated by commute peaks and weekend patterns reflecting recreational activities. Friday emerges as a transitional day, sharing the characteristics of both weekdays and weekends. Meanwhile, docking station usage rank distributions show remarkable consistency between weekdays and weekends for most cities, with London being a unique anomaly.
This study highlights the potential of BSS data to uncover urban mobility patterns and the underlying structures of cities. The results suggest that BSS usage reflects both intrinsic user behavior and external influences such as urban planning. 

}




\maketitle

\section{Introduction}
Cities represent one of the greatest inventions of humanity, serving as hubs where a large number of people communicate, exchange information and services, maintain complex socio-economic ecosystems and promote innovation \cite{glaeser2011triumph}. The defining characteristics of cities lie in their high population density and the proximity of individuals. To support efficient communication and smooth social activities, the physical transportation infrastructure within cities is indispensable. As urban areas expand and population density increases, the provision of a variety of mobility options becomes increasingly crucial for the maintenance of efficient economic and social activities. Moreover, a diverse set of transport modes within a city not only enhances movement efficiency, but also builds resilience in city transportation networks, thus contributing to long-term sustainability of the city \cite{Bettencourt2007-pr, Batty2008-mt}.

It is well known that human activities in urban areas display different spatial patterns and temporal rhythms \cite{Barbosa2015-ly,Franca2016,Yan2017-yt,Schlapfer2021-le}. Temporal rhythms emerge on multiple scales, such as seasonal, weekly, and daily, while spatial patterns are shaped by the geographic form of the city and the relative locations of commercial areas, entertainment facilities, and residential districts. Consequently, transportation systems must be implemented in ways that reflect these temporal and spatial patterns, and in turn, such systems reinforce and promote urban activities, creating a feedback loop that sustains city life.
Reflecting the importance of transportation systems, numerous studies have investigated a wide range of public transportation infrastructures, including airports \cite{Guimera2005-kn, Li2004-ya}, maritime transport \cite{Kaluza2010-st}, railways and subways \cite{Feng2017-xm, Louf2014-ee, Roth2011-kj, Chen2009-vg, Lee2008-ny, Li2007-qd}, and buses \cite{Wang2020-dc}. From a network science perspective, extensive research has focused on analyzing transportation networks. Broadly speaking, transportation systems can be divided into large-scale intercity networks and local intra-urban systems. Due to differences in their functions, each level exhibits different characteristics in the spatial distribution of travel density \cite{Liang2013-oy}.

Among the various forms of intra-urban public transportation, such as subways, buses, and taxis, new modes of shared mobility have emerged in recent years, driven by the rise of the sharing economy. This diversification of urban transportation options has brought attention to the Bike Sharing System (BSS), which provides lightweight and highly mobile transportation solutions. BSS programs can be broadly categorized into those with docking stations and those without. Users can access bicycles easily and spontaneously, making BSSs increasingly integral as a form of urban infrastructure that enhances the efficiency of intra-city movement. Recent studies have shown that the usage of BSS can reduce environmental burdens \cite{Ji2017-vw, Zhang2018-va} and improve public health \cite{Oja2011-xy}, thus contributing to sustainable urban development \cite{DeMaio2009-ft, Pucher2011-qt, Fishman2013-eo}.

A key advantage of BSSs is the availability of detailed operational data, including individual user attributes, usage times, trip durations, and docking station data. From a data-driven research perspective, BSSs serve as ideal models for studying urban transportation systems. Consequently, numerous studies have examined the spatio-temporal patterns and usage dynamics of BSSs, clarifying how these patterns mirror broader urban activity. In particular, weekday–weekend differences in urban rhythms manifest themselves directly in the temporal and spatial patterns of BSS use, prompting several research efforts in this area.
For example, Zhou et al. (2015) analyzed the ridership and movement patterns of the BSS in Chicago and identified distinct spatio-temporal differences between the use of weekday and weekend \cite{Zhou2015-ky}. Specifically, weekday usage peaked during commute hours, while weekend usage was highest during midday. By modeling bicycle flows from origins to destinations as networks and applying clustering techniques, they identified unique weekday and weekend clusters with different characteristics.
Kim et al. (2023) classified BSS usage in Seoul by pass types (daily, short-term, and long-term) and found that these differences reflect the underlying usage purposes \cite{Kim2023-pd}. Long-term passes tended to be associated with commuting, while short-term and daily passes were more likely to be tied to leisure activities. These distinctions in usage purpose were shown to directly influence user movement patterns, resulting in pronounced differences between weekdays and weekends.
Differences in weekday and weekend usage patterns also emerge in response to external factors. Basak et al. (2023) conducted a comparative analysis of the impact of the 2020 COVID-19 pandemic on BSS demand in 10 major US cities, revealing variations in how each city and each type of day (weekday vs. weekend) responded. Cities more supportive of bicycle use experienced greater impacts, and weekend ridership in particular decreased markedly, a trend attributed to the reduction in leisure-oriented trips during the pandemic \cite{Basak2023-gf}.

Building on this body of literature, which has established how differences in weekday and weekend usage patterns correlate with the underlying purpose and external factors, this study aims to identify both city-specific and universal characteristics of BSS usage patterns by comparing multiple major cities. We introduce quantitative measures to characterize usage differences in New York, London, Tokyo, Boston, Chicago, and Washington D.C. More specifically, we first compare the temporal patterns of daily BSS usage in these six cities. We then examine the spatial distributions of the docking stations used. While the former provides insight into temporal usage patterns, the latter illuminates spatial usage differences. By examining both dimensions, we seek to uncover commonalities and differences in BSS usage in diverse urban contexts.

\section{Data Description}\label{sec:data_detail}

In this section, we provide a detailed description of the datasets used in this comparative study of BSSs across six cities: “Citi Bike” in New York \cite{citibike_data}, “Santander Cycles” in London \cite{santander_data}, “Docomo Bikeshare” in Tokyo \cite{docomo_data}, “Bluebikes” in Boston \cite{bluebikes_data}, “Divvy” in Chicago \cite{divvy_data}, and “Capital Bikeshare” in Washington D.C.\ \cite{capital_data}. Previous research \cite{Gebhart2014-tk} has shown that variations in temperature can significantly affect the use of BSS. To control for this factor, we selected a continuous 30-day period between September and November 2023, during which the average temperature ranged from 13 ° C to 14 ° C \cite{meteostat} (Table~\ref{tab:data_detail}). All analyses in this study are based on data collected within that time frame.

It is important to note that the availability of data differs between two groups of cities: (a) New York, London, Boston, Chicago, and Washington D.C., and (b) Tokyo. For the first group (a), individual trip records are provided, which, despite minor differences in data formats between cities, always include the following: a unique trip ID, departure station ID, departure time, arrival station ID, arrival time, and the latitude and longitude of each station. In contrast, Tokyo (b) does not provide individual trip data, and no historical data set is available. Instead, an application programming interface (API) for the Tokyo BSS provides real-time information on each docking station, updated every minute. Consequently, we must infer user activity from minute-by-minute changes in the number of bicycles at each station.

Because we aim to compare all six cities, we transformed the trip-level data from the other five cities into minute-by-minute station-level usage data, consistent with the Tokyo dataset. Table~\ref{tab:data_detail} summarizes the key attributes of each city’s BSS. As shown, the total number of rentals and returns in Boston, Chicago, and Washington D.C.\ differs significantly from those in London, New York, and Tokyo. In particular, New York exhibits an order-of-magnitude higher volume of rentals compared to the first three cities. Tokyo also shows a larger gap between weekday and weekend usage than the other five cities, as reflected by the total number of rentals per station.

\begin{table}[ht]
  \centering
  \caption{Summary of BSS Data Across Six Cities}\label{tab:data_detail}
  \begin{tabular}{@{}lrrrrl@{}}
    \toprule
    \multicolumn{1}{c}{City} & \multicolumn{2}{c}{\# of Rentals/Returns} & \multicolumn{2}{c}{\# of Stations Used} & Data Collection \\
    \multicolumn{1}{c}{} & Weekdays & Weekends & Weekdays & Weekends & Period \\
    \midrule
    New York    & 5036538 & 1759118 & 2664  & 2594  & 2023-10-13 to 11-11 \\
    London      & 1210884 & 373178  & 803   & 802   & 2023-09-28 to 10-27 \\
    Tokyo       & 1563721 & 464659  & 1300  & 777   & 2023-11-01 to 11-30 \\
    Boston      & 528142  & 193625  & 468   & 468   & 2023-10-15 to 11-13 \\
    Chicago     & 464355  & 147714  & 1151  & 989   & 2023-11-01 to 11-30 \\
    Washington D.C.  & 549643  & 252463  & 750   & 747   & 2023-10-15 to 11-13 \\
    \bottomrule
  \end{tabular}
\end{table}

\section{Quantifying Differences in BSS Usage Patterns by Day of the Week}\label{sec:timeseries}

\begin{figure}[h]
\centering
\includegraphics[width=0.95\textwidth]{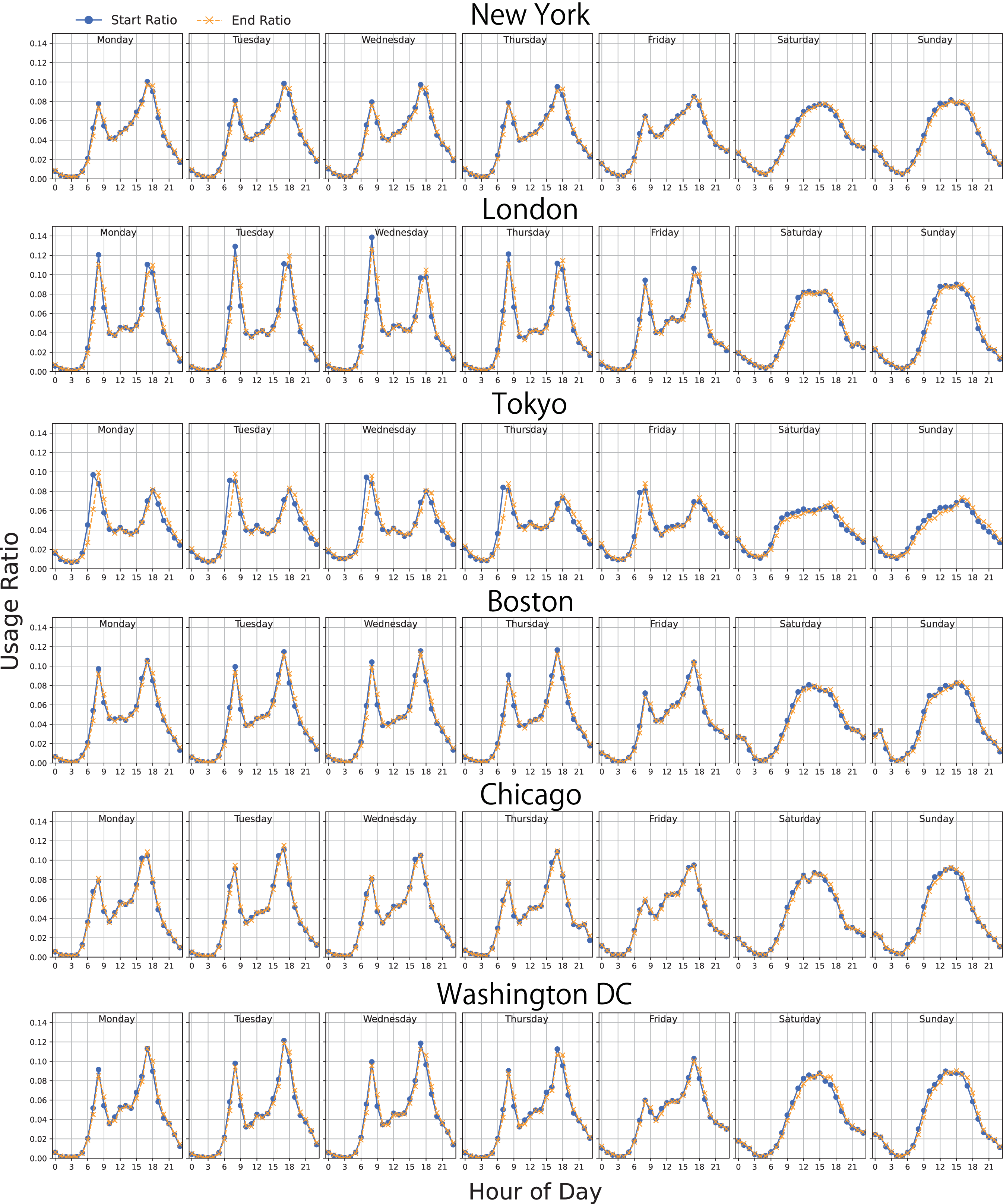}
\caption{Normalized probability distributions of BSS usage of six cities by day of the week. The horizontal axis denotes time (hour), with solid blue lines indicating rentals and dashed orange lines representing returns.}\label{fig:bss_week_fre}
\end{figure}

We first examine the differences in the usage patterns of BSS by day of the week, as highlighted in previous studies \cite{Zhou2015-ky, Loaiza-Monsalve2019-hu}. To facilitate comparisons between cities, we employed the normalized probability distribution of BSS usage, \( P(t) = N(t) / N_{\text{total}} \), where \( N(t) \) represents the frequency of use at time \( t \), and \( N_{\text{total}} \) is the total frequency of usage for each day of the week in a given city. Unlike the direct frequency distribution, this normalized approach allows us to compare usage patterns more effectively between cities with differing total levels of BSS activity.

Fig.~\ref{fig:bss_week_fre} shows the normalized probability distributions of the usage of BSS on weekdays and weekends in six cities. New York, London, Tokyo, Boston, Chicago and Washington D.C. In all cities, weekday usage shows a consistent bimodal pattern, with prominent peaks occurring during the morning and evening hours. These peaks, typically observed between 7:00 and 9:00 in the morning and between 17:00 and 19:00 in the evening, correspond to commute times, suggesting that BSSs are widely used as a commute option. In contrast, weekend usage exhibits a flatter distribution with less pronounced peaks, remaining relatively stable throughout the day. This indicates that BSSs are often employed for recreational or leisure activities on weekends, such as tourism or casual trips. Furthermore, all cities show minimal BSS activity during the late night and early morning hours (0:00 to 6:00), indicating that BSSs serve primarily as daytime modes of transportation. These patterns align with the findings of previous studies, which have demonstrated similar characteristics in the use of BSS in Chicago and other major urban areas \cite{Zhou2015-ky, Loaiza-Monsalve2019-hu}.

In addition to these shared patterns, each city exhibits unique characteristics in its usage distribution. In New York, the morning peak occurs earlier than in other cities, and the evening peak persists for a longer duration. This probably reflects the wide commute zones of the city and its vibrant nightlife. In London, the morning and evening peaks are particularly pronounced, with sharp contrasts between peak and off-peak hours. This suggests that the usage of BSS in London is heavily driven by commuting, underscoring the system’s role as a critical component of the city’s transportation network. In contrast, Tokyo shows a more balanced usage distribution throughout the day, with less pronounced morning and evening peaks compared to other cities. Tokyo also stands out for its relatively high nighttime usage, which can be attributed to the city’s active nightlife. Meanwhile, Boston, Chicago, and Washington D.C., exhibit similar patterns, with a bimodal weekday distribution characterized by a smaller morning peak and a larger evening peak.
These findings highlight both the universal features of BSS usage in cities and the localized variations shaped by each city's unique sociocultural and geographical context.

\section{Quantitative Analysis of BSS Usage Distributions Using Jensen-Shannon Divergence}\label{sec:JSD}

The results presented thus far indicate that the BSS usage probability distributions exhibit significant differences by day of the week, particularly between weekdays and weekends. In addition, the influence of city-specific characteristics and lifestyles is evident in the shapes of weekday distributions. To quantitatively evaluate these differences in BSS usage distributions, we employ Jensen-Shannon divergence (JSD), defined as follows. 

For two probability distributions \( P = [p_1, p_2, \dots, p_r] \) and \( Q = [q_1, q_2, \dots, q_r] \), and their mean distribution \( M = [ (p_1 + q_1)/2, (p_2 + q_2)/2, \dots, (p_r + q_r)/2 ] \), the JSD is given by:

\begin{equation}
D_{\text{JSD}}(P \parallel Q) = \frac{1}{2} D_{\text{KL}}(P \parallel M) + \frac{1}{2} D_{\text{KL}}(Q \parallel M), \label{eq:JSD}
\end{equation}
where \( D_{\text{KL}}(P \parallel Q) = \sum_{i=1}^r p_i \log({p_i}/{q_i}) \) is the Kullback-Leibler divergence (KLD). Unlike KLD, which is asymmetric (\( D_{\text{KL}}(P \parallel Q) \neq D_{\text{KL}}(Q \parallel P) \)), JSD is symmetric (\( D_{\text{JSD}}(P \parallel Q) = D_{\text{JSD}}(Q \parallel P) \)). The JSD value ranges from \( 0 \) to \( 1 \), with smaller values indicating greater similarity between distributions and larger values indicating greater differences. In this study, we used JSD to compare differences in normalized BSS usage probability distributions on days of the week.

\begin{figure}[h]
\centering
\includegraphics[width=0.98\textwidth]{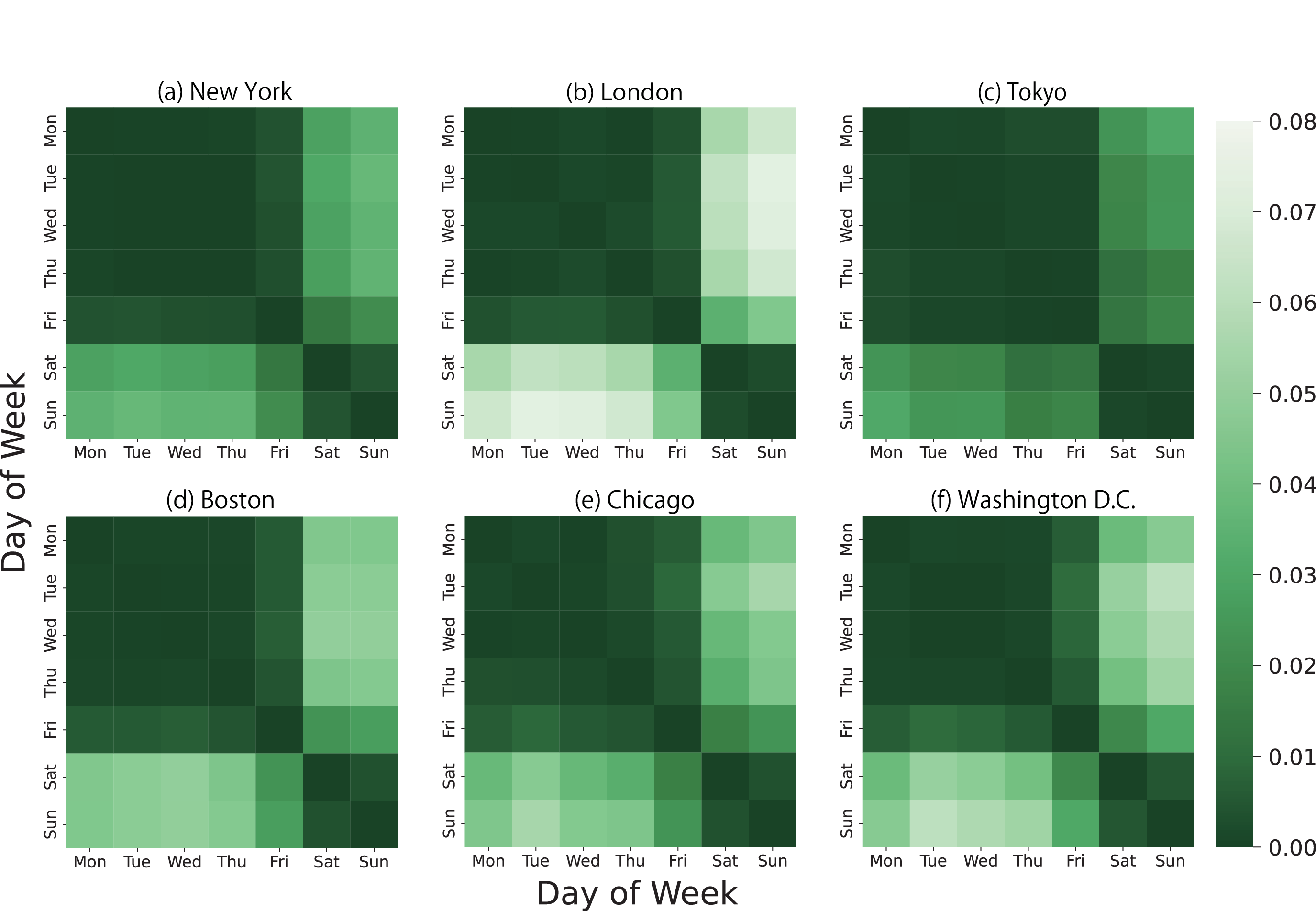}
\caption{Heatmaps of JSD values between days of the week for each city. Both axes represent days of the week (Monday to Sunday). The matrix elements denote the JSD values, color-coded on a green gradient scale ranging from [0.00, 0.08], where darker shades represent smaller values. Lower JSD values indicate greater similarity between distributions.}\label{fig:bss_week_heat}
\end{figure}

Fig.~\ref{fig:bss_week_heat} presents heatmaps of the JSD values for pairs of days of the week in the six cities. 
As inferred from the usage probability distributions in Fig.~\ref{fig:bss_week_fre}, there is a clear distinction between weekdays and weekends. Specifically, the JSD values for weekday pairs range from \( 0.000 \) to \( 0.010 \), and for weekend pairs from \( 0.000 \) to \( 0.005 \). In contrast, the JSD values between weekdays and weekends range from \( 0.012 \) to \( 0.073 \), indicating a clear separation.

Furthermore, Friday exhibits unique characteristics in all cities except Tokyo, with JSD values between Friday and weekends ranging from \( 0.013 \) to \( 0.045 \), compared to the range of \( 0.027 \) to \( 0.073 \) for other weekdays and weekends. This suggests that Friday serves as a transition day between weekday and weekend patterns, reflecting lifestyle trends in the use of BSS. Meanwhile, for Tokyo and New York, the JSD values between weekdays (excluding Friday) and weekends range from \( 0.012 \) to \( 0.038 \), smaller than those of other cities (\( 0.033 \) to \( 0.073 \)). This finding implies that Tokyo and New York exhibit less pronounced distinctions between weekday and weekend BSS usage compared to other cities.

\section{JSD Network Analysis for Comparing BSS Usage Across Cities}\label{sec:JSDnetwork}

\begin{figure}[h]
\centering
\includegraphics[width=0.98\textwidth]{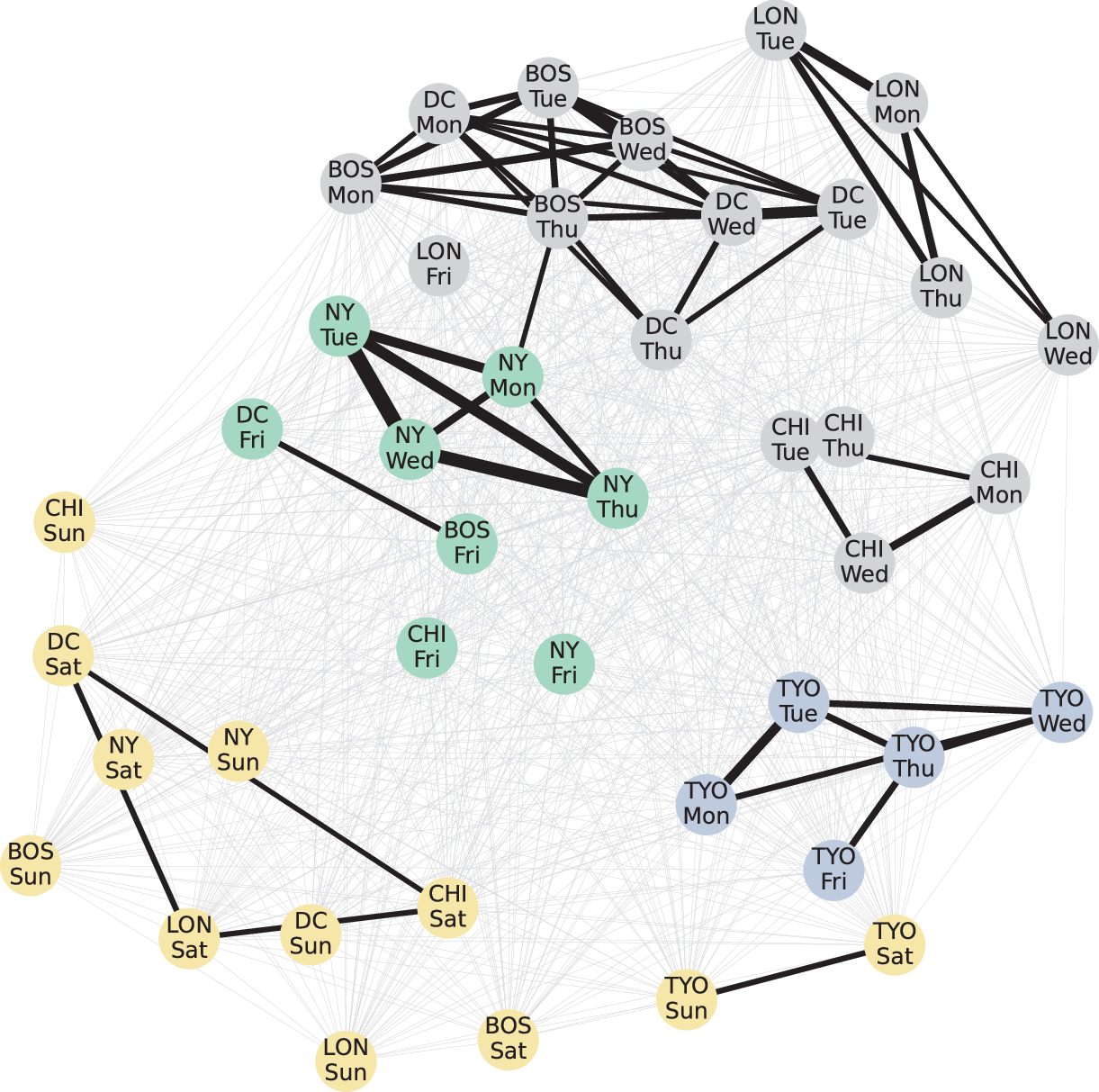}
\caption{JSD network representing relationships between city-day combinations. Each node corresponds to a specific city-day pair, and edges represent the inverse of the JSD values as weights in a weighted network. The top 50 edges with the largest weights are shown with thickness proportional to their values, while lower-ranked edges are displayed as thin, faint lines. Node colors indicate clusters identified using the Louvain method, and node positions are determined using the spring layout.}\label{fig:jsd_net}
\end{figure}

To further explore the differences between cities, we constructed a JSD network to compare BSS usage distributions across all pairs of city days. With six cities and seven days per city, the analysis involves 42 distributions and \( 42 \times 42 = 1764 \) JSD relationships. Visualizing such a large number of relationships as a heatmap is impractical, so we represent these relationships as a JSD network (Fig.~\ref{fig:jsd_net}). The nodes represent pairs of city days, and the edge weights are given by the inverse of the JSD values, forming a complete weighted network. The Louvain method \cite{Blondel2008} is used to detect communities based on edge weights.

The Louvain method identifies four main communities. The first community (yellow) represents weekends in all cities. The second (blue) comprises the weekdays in Tokyo. The third (green) includes New York weekdays along with Fridays in Chicago, Washington D.C., and Boston. Finally, the fourth (gray) consists of weekdays (excluding Fridays) for London, Chicago, Washington D.C., and Boston, with London uniquely including Friday as well. The spatial arrangement of the nodes in the spring layout reflects the similarity of the BSS usage patterns, with the weekend nodes grouped distinctly and the weekday nodes exhibiting closer connections based on similarities between cities and between days.

The network visualzation reveals that the usage of BSS on weekends shows minimal variation between cities, while the usage on weekends exhibits more pronounced differences between cities. Tokyo’s weekday cluster is distinctly separate, indicating unique weekday usage patterns compared to other cities. Similarly, New York’s weekdays form a tightly connected cluster, highlighting the uniformity of weekday BSS usage. Meanwhile, cities such as Chicago, Washington D.C., and Boston display nuanced differences within their weekday patterns, as indicated by clustering in the community.
These findings underscore the effectiveness of JSD network analysis in capturing variations between cities and between days in BSS usage distributions.

\section{Comparison of Rank Distributions of Docking Station Usage}\label{sec:rank_distribution}

\begin{figure}[h]
  \centering
  \includegraphics[width=0.98\textwidth]{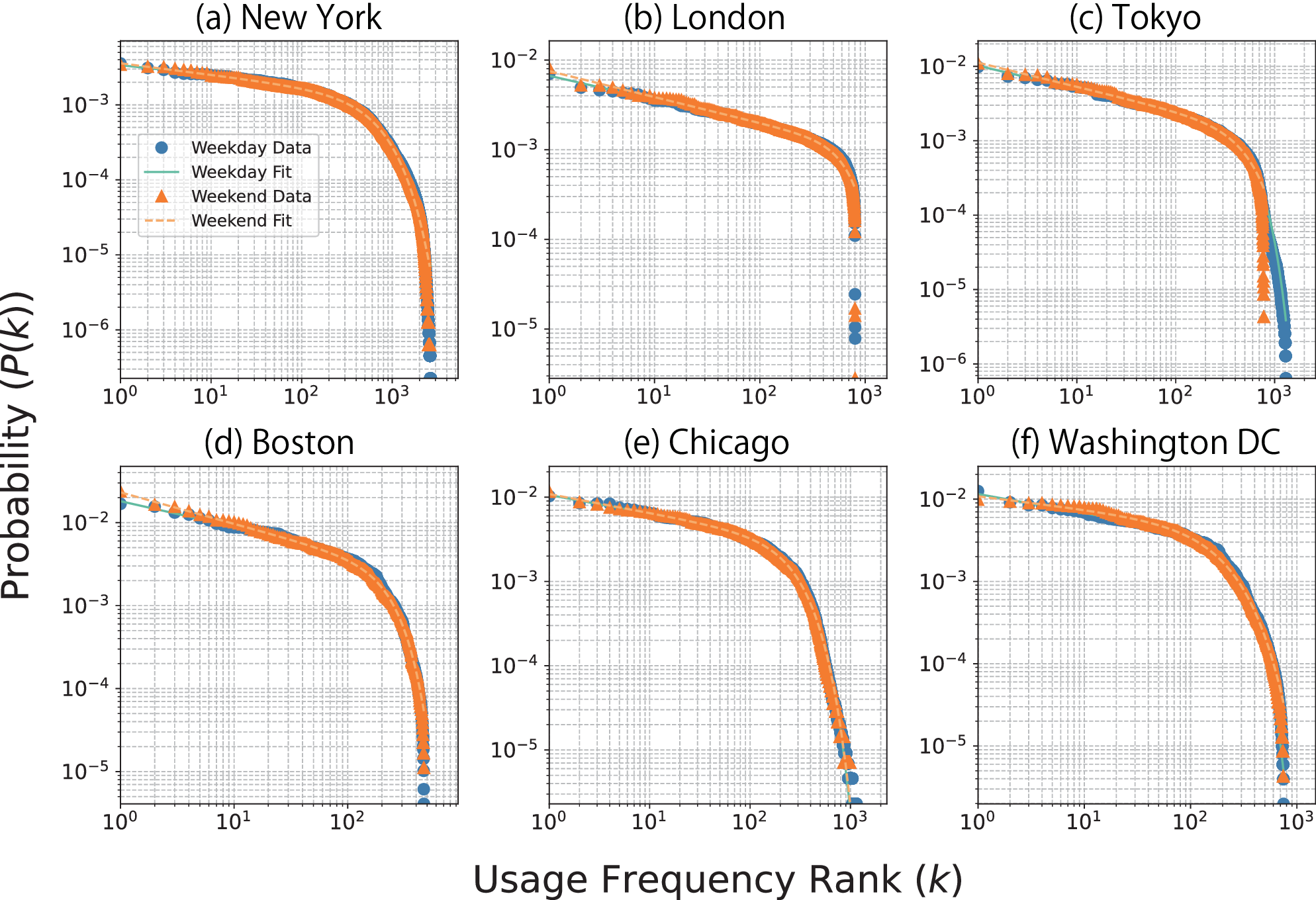}
  \caption{Comparison of docking station usage rank distributions on weekdays and weekends across cities. Weekday data are shown as blue circles, while weekend data are represented as orange triangles. Dashed lines (green) indicate fitted curves for weekday and weekend data. The \( x \)-axis represents the rank of docking station usage, and the \( y \)-axis represents the normalized usage frequency, both on a log-log scale.}\label{fig:rank_dist}
\end{figure}

In the previous chapters, we focused on the temporal characteristics of BSS usage, quantitatively evaluating the patterns across different days of the week and cities. The results revealed that BSS usage is influenced by the lifestyles of urban residents, which are largely independent of the specific city. On weekdays, BSS usage is primarily driven by commutes to business areas, while on weekends it is associated with leisure activities such as tourism and recreation. This implies that the destinations, and therefore the docking stations, used by commuters on weekdays differ from those used by recreational users on weekends. 

To explore this further, this chapter examines the rank distributions of docking station usage frequencies, analyzing how these ranks differ between weekdays and weekends, and investigating the changes in docking station rankings over time. The rank frequency distribution of docking station usage in transportation systems has been widely studied, revealing characteristic patterns such as power laws and truncated exponential distributions \cite{Kaluza2010-st, Lee2008-ny, Li2004-ya, Roth2011-kj}. Recent findings also suggest that frequency distributions differ between inter-city and intra-city transportation systems, with inter-city systems exhibiting power-law distributions and intra-city systems following exponential distributions \cite{Liang2013-oy}.
\begin{table}[]
  \caption{Fitting Parameters for the Function in Equation~(\ref{eq:fitting_func})} 
  \centering
  \begin{tabular}{@{}lrrrrrr@{}}
  \toprule
   City & \multicolumn{2}{c}{$\alpha$} & \multicolumn{2}{c}{$\beta$} & \multicolumn{2}{c}{$\gamma$} \\
        & Weekday & Weekend & Weekday & Weekend & Weekday & Weekend \\
   \midrule
   New York    & $0.129$ & $0.153$ & $5.6 \times 10^{-4}$ & $2.5 \times 10^{-4}$ & $1.16$ & $1.26$ \\
   London      & $0.267$ & $0.289$ & $3.6 \times 10^{-8}$ & $4.4 \times 10^{-8}$ & $2.57$ & $2.54$ \\
   Tokyo       & $0.314$ & $0.324$ & $2.2 \times 10^{-6}$ & $2.1 \times 10^{-5}$ & $2.06$ & $1.70$ \\
   Boston      & $0.305$ & $0.379$ & $2.0 \times 10^{-5} $ & $1.1 \times 10^{-5}$ & $1.99$ & $2.08$ \\
   Chicago     & $0.221$ & $0.224$ & $1.0 \times 10^{-4}$ & $1.3 \times 10^{-4}$ & $1.62$ & $1.57$ \\
   Washington D.C.  & $0.230$ & $0.157$ & $3.7 \times 10^{-5}$ & $2.0 \times 10^{-3}$ & $1.82$ & $1.18$ \\
   \bottomrule
  \end{tabular} \label{table:fitting_parametor}
\end{table}

Fig.~\ref{fig:rank_dist} presents the rank distributions of the usage of the docking station on weekdays (blue circles) and weekends (orange triangles) in cities. Surprisingly, the distributions for weekdays and weekends appear to be remarkably similar across all cities. To quantify these patterns, we fitted the rank distributions for both weekdays and weekends using the function:

\begin{equation}
  P(k) \propto k^{-\alpha}\exp(-\beta k^{\gamma}), \label{eq:fitting_func}
\end{equation}
where \( P(k) \) represents the normalized usage frequency of docking stations ranked \( k \), and \( \alpha, \beta, \gamma \) are fitting parameters. This function combines a power-law term and an exponential cut-off, capturing the characteristics of the distributions for both high-ranking and low-ranking stations. The key parameters shaping the distribution are \( \alpha \), the power-law exponent, and \( \gamma \), which determines the exponential decay.

The fitted curves for weekdays (dashed lines) and weekends (dotted lines) are shown in Fig.~\ref{fig:rank_dist}, with the corresponding parameter values listed in Table~\ref{table:fitting_parametor}. The fitted function accurately reproduces the observed data for all cities both on weekdays and weekends. In particular, the high-rank portions of the distributions follow power laws, with similar values for the exponent \( \alpha \) on weekdays and weekends. Furthermore, the lower-rank portions exhibit consistent values for \( \gamma \), indicating similar patterns of exponential decay.

These results are unexpected given the differences in weekday and weekend usage purposes identified in previous sections. To better understand these similarities between weekdays and weekends, we compare the usage rankings of individual docking stations between weekdays and weekends to assess the extent of their divergence.


\begin{figure}[h]
  \includegraphics[width=0.98\textwidth]{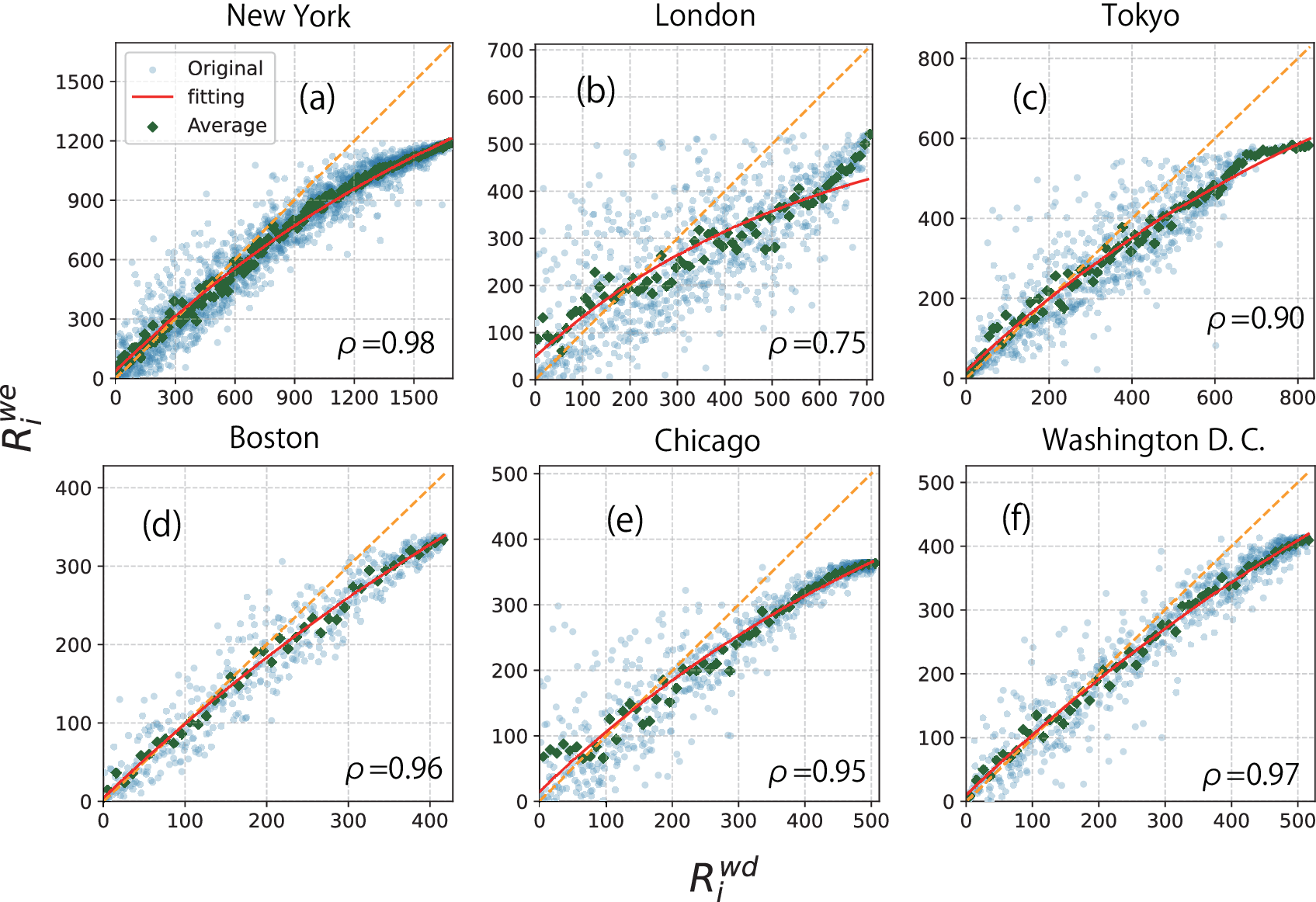}
  \caption{Comparison of docking station usage rankings between weekdays and weekends. The \( x \)-axis represents weekday rankings, and the \( y \)-axis represents weekend rankings. Panels (a)–(f) correspond to New York, London, Tokyo, Boston, Chicago, and Washington D.C., respectively. Circular points (light blue) represent the rankings of docking stations based on their usage frequency. Diamond points (green) indicate the averaged weekend rankings for every 10 weekday ranking bins. The solid red line shows the fitted curve, while the orange dashed line represents the \( y = x \) guideline, where equal rankings on weekdays and weekends would align. Spearman's rank correlation coefficient (\( \rho \)) values are displayed for each panel to quantify the strength of the correlation.}\label{fig:rank_wd_vs_we}
\end{figure}

Fig.~\ref{fig:rank_wd_vs_we} compares the usage rankings of the docking stations between weekdays and weekends. 
The rankings were assigned according to the frequency of use, limited to the docking stations used at least once on both weekdays and weekends.However, it is important to note that the weekday and weekend rankings reflect the original data, which include docking stations that may not have been used on both days. As a result, the total number of rankings differs between weekdays and weekends. The \( x \)-axis represents the weekly rankings and the \( y \)-axis represents the weekend rankings. 
Each blue point corresponds to a docking station. Additionally, the original data were divided into bins of 10 weekday ranks each, and the average weekend rank for each bin was calculated and plotted (green diamond points). The points located on the orange dashed line (\( y = x \)) indicate equal ranking on weekdays and weekends.The solid red line represents a fitted curve using the function:
\begin{equation}
  y(x) = \left( b - \frac{1}{a} \right) \left( 1 - a \right)^x + \frac{1}{a}, \label{eq:fitting_func2}
\end{equation}
where \( a, b \) are the fitting parameters. This function was derived using a simplified model, detailed in the Appendix~\ref{app:secA1}. The values of \( a \), \( b \) and the Root Mean Squared Error (RMSE) for each city are listed in Table~\ref{table:param_a_b}.

\begin{table}[]
  \caption{Fitting parameters \( a \), \( b \), and RMSE for the function in Equation~(\ref{eq:fitting_func2}).}
  \centering
  \begin{tabular}{@{}lrrr@{}}
  \toprule
   City & $a$ & $b$ & RMSE\\
   \midrule
   New York    & $4.38 \times 10^4$ & $38.7$ & $76.7$ \\
   London      & $1.70 \times 10^3$ & $49.1$ & $92.1$ \\
   Tokyo       & $8.72 \times 10^4$ & $19.1$ & $70.6$ \\
   Boston      & $1.08 \times 10^3$ & $4.5$ & $27.3$\\
   Chicago     & $1.43 \times 10^3$ & $14.6$ & $36.1$ \\
   Washington  & $1.20 \times 10^3$ & $10.1$ & $33.4$ \\
   \bottomrule
  \end{tabular} \label{table:param_a_b}
\end{table}

In Fig.~\ref{fig:rank_wd_vs_we}, clear differences emerge between New York (a), Tokyo (c), Boston (d), Chicago (e), and Washington D.C.\ (f) compared to London (b). For the first group of cities, despite some variability, there is a strong correlation between the weekday and weekend rankings of docking station use, with Spearman's rank correlation coefficient \( \rho \geq 0.90 \). In particular, for higher-ranked docking stations, the data points align closely with the \( y = x \) line, indicating similar rankings for heavily used docking stations on both weekdays and weekends. As the ranks increase, the data points deviate upward, showing a convex relationship. 
In contrast, London (b) exhibits greater variability in the data points and a weaker correlation, with a Spearman's rank correlation coefficient \( \rho = 0.75 \). Furthermore, the convex trend to the top in the other cities is absent in London, making it difficult to achieve a good fit using Equation~(\ref{eq:fitting_func2}).

These findings highlight the varying degrees of correlation between the use of docking stations on weekdays and weekends in cities. Although most cities demonstrate similar patterns in high-frequency station usage, London appears to deviate, suggesting distinct dynamics in its weekday and weekend docking station utilization.

\section{Conclusions}\label{sec:discussion}

This study analyzed bike sharing systems (BSS) usage patterns in six cities, New York, London, Tokyo, Boston, Chicago, and Washington D.C., over a continuous 30-day period between September and November 2023, during which average temperatures were comparable. Using trip data and docking station data, we examined both the temporal and spatial characteristics of BSS usage, providing a detailed comparison between cities and days of the week.

The temporal analysis, using the Jensen-Shannon divergence (JSD), revealed significant differences between weekday and weekend usage patterns in all cities. Among weekdays, Friday often exhibited transitional characteristics, blending features of both weekdays and weekends. This observation is consistent with previous research \cite{Zhou2015-ky, Loaiza-Monsalve2019-hu}, which highlighted the influence of commuting during the week and recreational activities during the weekends. By constructing a JSD-based network and performing community detection, we demonstrated that weekday activities are more closely related to city-specific rhythms, while weekend activities are less dependent on these factors, strengthening findings from earlier studies \cite{Basak2023-gf}.
In the ranking analysis, the rank distributions of docking station usage were remarkably consistent on weekdays and weekends within each city. However, significant differences emerged when comparing the ranking of docking stations on weekday and weekends in cities. Cities such as New York, Tokyo, Boston, Chicago, and Washington D.C.\ showed strong correlations, while London exhibited weaker correlations, indicating distinct patterns of usage. These results extend previous studies exploring the utilization of docking stations by emphasizing how consistent rank distributions are shaped by urban planning and cultural factors.

A limitation of this study is the restricted 30-day analysis period, driven by data constraints in Tokyo. Expanding the data set to cover longer timescales would allow us to examine seasonal and long-term changes in BSS usage. This could provide deeper insight into city-specific differences and universal behaviors in BSS usage. Future research could also investigate the relationship between docking station locations and surrounding facilities to better understand the factors influencing docking station usage patterns in cities.
By combining longer-term data and additional urban metrics, such as population density, public transit connectivity, and land use, future studies could further elucidate how BSSs interact with broader urban systems. These insights would support the development of sustainable and user-centered urban mobility solutions.







\section*{Declarations}

\begin{itemize}
\item \textbf{Availability of data and materials}

The data used in this study are open access and were obtained from the official websites of Citi Bike (\url{https://citibikenyc.com/system-data}), Santander Cycles (\url{https://cycling.data.tfl.gov.uk/}), Bluebikes (\url{https://bluebikes.com/system-data}), Divvy (\url{https://divvybikes.com/system-data}), and Capital Bikeshare (\url{https://capitalbikeshare.com/system-data}). Furthermore, data from Docomo Bikeshare (\url{https://ckan.odpt.org/dataset/c_bikeshare_gbfs-d-bikeshare}) were obtained using an API provided by NTT Docomo. Requests for data sharing will be considered by the authors upon reasonable request.

\item \textbf{Competing interests}

Not applicable.

\item \textbf{Funding}

SK and YB received research funding from Musashino University.

\item \textbf{Authors' contributions}

SK conceptualized the overall study design, provided ideas, performed data analysis and interpretation, and drafted the manuscript. YB acquired data using APIs and conducted portions of the data analysis. HS contributed ideas for data analysis, provided guidance on the structure of the manuscript, and revised the manuscript. All authors reviewed and approved the final version of the manuscript.

\item \textbf{Acknowledgements}

S.K. thanks Musashino University for granting a sabbatical that enabled focused research. This paper was written using ChatGPT 4o, but its usage was limited to language translation only. The content of the paper is entirely original, written solely by the authors, and the final English version was thoroughly reviewed and manually edited by the authors.

\end{itemize}


\begin{appendices}
\section{Derivation of the Fitting Function in Equation~(\ref{eq:fitting_func2})}\label{app:secA1}

Here, we derive a simplified mathematical model to describe the statistical characteristics of the strong correlation observed between weekday and weekend docking station  usage rankings, as shown in Fig.~\ref{fig:rank_wd_vs_we}. This analysis focuses on cities such as New York (a), Tokyo (c), Boston (d), Chicago (e) and Washington D.C.\ (f), where the rankings exibit a strong correlation. In these cities, the data points generally align with \( y = x \) at higher ranks, while showing an upward convex trend at lower ranks.  

Let \( k \) represent the weekday rank and \( \langle S(k) \rangle \) denote the average weekend rank corresponding to \( k \). We assume \( \langle S(k) \rangle \) is a monotonically increasing with \( k \), reflecting the strong correlation. Let \( k = \{1, 2, \dots, N\} \), \( 1 < \langle S(k) \rangle < M \), and \( N > M \), where \( M \) is the maximum weekend rank. To determine the relationship between \( k \) and \( \langle S(k) \rangle \), we iteratively compute \( \langle S(k) \rangle \) starting from the highest weekday rank (\( k = 1 \)).

For the initial rank (\( k = 1 \)), we set \( \langle S(1) \rangle = S_1 \) as the initial average weekend rank. For the second rank (\( k = 2 \)), the increment in the average weekend rank is denoted as \( \Delta \langle S(1) \rangle \). Given the strong correlation, the increase from \( k = 1 \) to \( k = 2 \) would ideally be \( \Delta \langle S(1) \rangle = 1 \). However, since \( N > M \), some docking stations are used only on weekdays and may not appear in the weekend rankings. To account for this, we introduce a parameter \( 0 < a < 1 \) to represent the proportional likelihood that a docking station ranked on weekdays is not included in the weekend rankings. Assuming this likelihood is proportional to the rank \( \langle S(1) \rangle \), the probability of a docking station appearing in the weekend rankings is given by \( 1 - a\langle S(1) \rangle \). Thus, the increment becomes \( \Delta \langle S(1) \rangle = (1 - a\langle S(1) \rangle) \times 1 \). Consequently, we have:

\begin{equation}
\langle S(2) \rangle = \langle S(1) \rangle + \Delta \langle S(1) \rangle = \langle S(1) \rangle + 1 - a\langle S(1) \rangle. \label{eq:rank_relation_2}
\end{equation}

Similarly, for the \( k \)-th weekday rank, the corresponding weekend rank is given by:

\begin{equation}
\langle S(k) \rangle = \langle S(k-1) \rangle + 1 - a \langle S(k-1) \rangle. \label{eq:rank_relation_k}
\end{equation}

By solving this recurrence relation, we obtain:

\begin{equation}
\langle S(k) \rangle = \left( S_1 - \frac{1}{a} \right) (1 - a)^k + \frac{1}{a}, \label{eq:rank_relation}
\end{equation}
which corresponds to Equation~(\ref{eq:fitting_func2}). This model is a simplified approximation based on the assumption of \( N > M \) and strong correlation between weekday and weekend rankings. Despite its simplicity, the derived function fits the data in Fig.~\ref{fig:rank_wd_vs_we} well, capturing the key features of the observed correlations.

\end{appendices}


\ifx \bisbn   \undefined \def \bisbn  #1{ISBN #1}\fi
\ifx \binits  \undefined \def \binits#1{#1}\fi
\ifx \bauthor  \undefined \def \bauthor#1{#1}\fi
\ifx \batitle  \undefined \def \batitle#1{#1}\fi
\ifx \bjtitle  \undefined \def \bjtitle#1{#1}\fi
\ifx \bvolume  \undefined \def \bvolume#1{\textbf{#1}}\fi
\ifx \byear  \undefined \def \byear#1{#1}\fi
\ifx \bissue  \undefined \def \bissue#1{#1}\fi
\ifx \bfpage  \undefined \def \bfpage#1{#1}\fi
\ifx \blpage  \undefined \def \blpage #1{#1}\fi
\ifx \burl  \undefined \def \burl#1{\textsf{#1}}\fi
\ifx \doiurl  \undefined \def \doiurl#1{\url{https://doi.org/#1}}\fi
\ifx \betal  \undefined \def \betal{\textit{et al.}}\fi
\ifx \binstitute  \undefined \def \binstitute#1{#1}\fi
\ifx \binstitutionaled  \undefined \def \binstitutionaled#1{#1}\fi
\ifx \bctitle  \undefined \def \bctitle#1{#1}\fi
\ifx \beditor  \undefined \def \beditor#1{#1}\fi
\ifx \bpublisher  \undefined \def \bpublisher#1{#1}\fi
\ifx \bbtitle  \undefined \def \bbtitle#1{#1}\fi
\ifx \bedition  \undefined \def \bedition#1{#1}\fi
\ifx \bseriesno  \undefined \def \bseriesno#1{#1}\fi
\ifx \blocation  \undefined \def \blocation#1{#1}\fi
\ifx \bsertitle  \undefined \def \bsertitle#1{#1}\fi
\ifx \bsnm \undefined \def \bsnm#1{#1}\fi
\ifx \bsuffix \undefined \def \bsuffix#1{#1}\fi
\ifx \bparticle \undefined \def \bparticle#1{#1}\fi
\ifx \barticle \undefined \def \barticle#1{#1}\fi
\bibcommenthead
\ifx \bconfdate \undefined \def \bconfdate #1{#1}\fi
\ifx \botherref \undefined \def \botherref #1{#1}\fi
\ifx \url \undefined \def \url#1{\textsf{#1}}\fi
\ifx \bchapter \undefined \def \bchapter#1{#1}\fi
\ifx \bbook \undefined \def \bbook#1{#1}\fi
\ifx \bcomment \undefined \def \bcomment#1{#1}\fi
\ifx \oauthor \undefined \def \oauthor#1{#1}\fi
\ifx \citeauthoryear \undefined \def \citeauthoryear#1{#1}\fi
\ifx \endbibitem  \undefined \def \endbibitem {}\fi
\ifx \bconflocation  \undefined \def \bconflocation#1{#1}\fi
\ifx \arxivurl  \undefined \def \arxivurl#1{\textsf{#1}}\fi
\csname PreBibitemsHook\endcsname

\begin{itemize}
\bibitem[\protect\citeauthoryear{Glaeser}{2011}]{glaeser2011triumph}
\begin{bbook}
\bauthor{\bsnm{Glaeser}, \binits{E.}}:
\bbtitle{Triumph of the City: How Our Greatest Invention Makes Us Richer, Smarter, Greener, Healthier, and Happier}.
\bpublisher{Penguin Press},
\blocation{New York}
(\byear{2011}).
\burl{https://www.amazon.com/Triumph-City-Greatest-Invention-Healthier-ebook/dp/B0049U4HTW}
\end{bbook}
\endbibitem

\bibitem[\protect\citeauthoryear{Bettencourt et~al.}{2007}]{Bettencourt2007-pr}
\begin{barticle}
\bauthor{\bsnm{Bettencourt}, \binits{L.M.A.}},
\bauthor{\bsnm{Lobo}, \binits{J.}},
\bauthor{\bsnm{Helbing}, \binits{D.}},
\bauthor{\bsnm{Kühnert}, \binits{C.}},
\bauthor{\bsnm{West}, \binits{G.B.}}:
\batitle{Growth, innovation, scaling, and the pace of life in cities}.
\bjtitle{Proc. Natl. Acad. Sci. U. S. A.}
\bvolume{104}(\bissue{17}),
\bfpage{7301}--\blpage{7306}
(\byear{2007})
\end{barticle}
\endbibitem

\bibitem[\protect\citeauthoryear{Batty}{2008}]{Batty2008-mt}
\begin{barticle}
\bauthor{\bsnm{Batty}, \binits{M.}}:
\batitle{The size, scale, and shape of cities}.
\bjtitle{Science}
\bvolume{319}(\bissue{5864}),
\bfpage{769}--\blpage{771}
(\byear{2008})
\end{barticle}
\endbibitem

\bibitem[\protect\citeauthoryear{Barbosa et~al.}{2015}]{Barbosa2015-ly}
\begin{barticle}
\bauthor{\bsnm{Barbosa}, \binits{H.}},
\bauthor{\bsnm{Lima-Neto}, \binits{F.B.}},
\bauthor{\bsnm{Evsukoff}, \binits{A.}},
\bauthor{\bsnm{Menezes}, \binits{R.}}:
\batitle{The effect of recency to human mobility}.
\bjtitle{EPJ Data Sci.}
\bvolume{4}(\bissue{1}),
\bfpage{1}--\blpage{14}
(\byear{2015})
\end{barticle}
\endbibitem

\bibitem[\protect\citeauthoryear{Fran{\c{c}}a et~al.}{2016}]{Franca2016}
\begin{barticle}
\bauthor{\bsnm{Fran{\c{c}}a}, \binits{U.}},
\bauthor{\bsnm{Sayama}, \binits{H.}},
\bauthor{\bsnm{McSwiggen}, \binits{C.}},
\bauthor{\bsnm{Daneshvar}, \binits{R.}},
\bauthor{\bsnm{Bar-Yam}, \binits{Y.}}:
\batitle{Visualizing the “heartbeat” of a city with tweets}.
\bjtitle{Complexity}
\bvolume{21}(\bissue{6}),
\bfpage{280}--\blpage{287}
(\byear{2016})
\end{barticle}
\endbibitem

\bibitem[\protect\citeauthoryear{Yan et~al.}{2017}]{Yan2017-yt}
\begin{barticle}
\bauthor{\bsnm{Yan}, \binits{X.-Y.}},
\bauthor{\bsnm{Wang}, \binits{W.-X.}},
\bauthor{\bsnm{Gao}, \binits{Z.-Y.}},
\bauthor{\bsnm{Lai}, \binits{Y.-C.}}:
\batitle{Universal model of individual and population mobility on diverse spatial scales}.
\bjtitle{Nat. Commun.}
\bvolume{8}(\bissue{1}),
\bfpage{1639}
(\byear{2017})
\end{barticle}
\endbibitem

\bibitem[\protect\citeauthoryear{Schläpfer et~al.}{2021}]{Schlapfer2021-le}
\begin{barticle}
\bauthor{\bsnm{Schläpfer}, \binits{M.}},
\bauthor{\bsnm{Dong}, \binits{L.}},
\bauthor{\bsnm{O'Keeffe}, \binits{K.}},
\bauthor{\bsnm{Santi}, \binits{P.}},
\bauthor{\bsnm{Szell}, \binits{M.}},
\bauthor{\bsnm{Salat}, \binits{H.}},
\bauthor{\bsnm{Anklesaria}, \binits{S.}},
\bauthor{\bsnm{Vazifeh}, \binits{M.}},
\bauthor{\bsnm{Ratti}, \binits{C.}},
\bauthor{\bsnm{West}, \binits{G.B.}}:
\batitle{The universal visitation law of human mobility}.
\bjtitle{Nature}
\bvolume{593}(\bissue{7860}),
\bfpage{522}--\blpage{527}
(\byear{2021})
\end{barticle}
\endbibitem

\bibitem[\protect\citeauthoryear{Guimerà et~al.}{2005}]{Guimera2005-kn}
\begin{barticle}
\bauthor{\bsnm{Guimerà}, \binits{R.}},
\bauthor{\bsnm{Mossa}, \binits{S.}},
\bauthor{\bsnm{Turtschi}, \binits{A.}},
\bauthor{\bsnm{Amaral}, \binits{L.A.N.}}:
\batitle{The worldwide air transportation network: Anomalous centrality, community structure, and cities' global roles}.
\bjtitle{Proc. Natl. Acad. Sci. U. S. A.}
\bvolume{102}(\bissue{22}),
\bfpage{7794}--\blpage{7799}
(\byear{2005})
\end{barticle}
\endbibitem

\bibitem[\protect\citeauthoryear{Li and Cai}{2004}]{Li2004-ya}
\begin{barticle}
\bauthor{\bsnm{Li}, \binits{W.}},
\bauthor{\bsnm{Cai}, \binits{X.}}:
\batitle{Statistical analysis of airport network of china}.
\bjtitle{Phys. Rev. E}
\bvolume{69}(\bissue{4}),
\bfpage{046106}
(\byear{2004})
\end{barticle}
\endbibitem

\bibitem[\protect\citeauthoryear{Kaluza et~al.}{2010}]{Kaluza2010-st}
\begin{barticle}
\bauthor{\bsnm{Kaluza}, \binits{P.}},
\bauthor{\bsnm{Kölzsch}, \binits{A.}},
\bauthor{\bsnm{Gastner}, \binits{M.T.}},
\bauthor{\bsnm{Blasius}, \binits{B.}}:
\batitle{The complex network of global cargo ship movements}.
\bjtitle{J. R. Soc. Interface}
\bvolume{7}(\bissue{48}),
\bfpage{1093}--\blpage{1103}
(\byear{2010})
\end{barticle}
\endbibitem

\bibitem[\protect\citeauthoryear{Feng et~al.}{2017}]{Feng2017-xm}
\begin{barticle}
\bauthor{\bsnm{Feng}, \binits{J.}},
\bauthor{\bsnm{Li}, \binits{X.}},
\bauthor{\bsnm{Mao}, \binits{B.}},
\bauthor{\bsnm{Xu}, \binits{Q.}},
\bauthor{\bsnm{Bai}, \binits{Y.}}:
\batitle{Weighted complex network analysis of the beijing subway system: Train and passenger flows}.
\bjtitle{Physica A}
\bvolume{474},
\bfpage{213}--\blpage{223}
(\byear{2017})
\end{barticle}
\endbibitem

\bibitem[\protect\citeauthoryear{Louf et~al.}{2014}]{Louf2014-ee}
\begin{barticle}
\bauthor{\bsnm{Louf}, \binits{R.}},
\bauthor{\bsnm{Roth}, \binits{C.}},
\bauthor{\bsnm{Barthelemy}, \binits{M.}}:
\batitle{Scaling in transportation networks}.
\bjtitle{PLoS One}
\bvolume{9}(\bissue{7}),
\bfpage{102007}
(\byear{2014})
\end{barticle}
\endbibitem

\bibitem[\protect\citeauthoryear{Roth et~al.}{2011}]{Roth2011-kj}
\begin{barticle}
\bauthor{\bsnm{Roth}, \binits{C.}},
\bauthor{\bsnm{Kang}, \binits{S.M.}},
\bauthor{\bsnm{Batty}, \binits{M.}},
\bauthor{\bsnm{Barthélemy}, \binits{M.}}:
\batitle{Structure of urban movements: polycentric activity and entangled hierarchical flows}.
\bjtitle{PLoS One}
\bvolume{6}(\bissue{1}),
\bfpage{15923}
(\byear{2011})
\end{barticle}
\endbibitem

\bibitem[\protect\citeauthoryear{Chen et~al.}{2009}]{Chen2009-vg}
\begin{barticle}
\bauthor{\bsnm{Chen}, \binits{C.}},
\bauthor{\bsnm{Chen}, \binits{J.}},
\bauthor{\bsnm{Barry}, \binits{J.}}:
\batitle{Diurnal pattern of transit ridership: a case study of the new york city subway system}.
\bjtitle{J. Transp. Geogr.}
\bvolume{17}(\bissue{3}),
\bfpage{176}--\blpage{186}
(\byear{2009})
\end{barticle}
\endbibitem

\bibitem[\protect\citeauthoryear{Lee et~al.}{2008}]{Lee2008-ny}
\begin{barticle}
\bauthor{\bsnm{Lee}, \binits{K.}},
\bauthor{\bsnm{Jung}, \binits{W.-S.}},
\bauthor{\bsnm{Park}, \binits{J.S.}},
\bauthor{\bsnm{Choi}, \binits{M.Y.}}:
\batitle{Statistical analysis of the metropolitan seoul subway system: Network structure and passenger flows}.
\bjtitle{Physica A}
\bvolume{387}(\bissue{24}),
\bfpage{6231}--\blpage{6234}
(\byear{2008})
\end{barticle}
\endbibitem

\bibitem[\protect\citeauthoryear{Li and Cai}{2007}]{Li2007-qd}
\begin{barticle}
\bauthor{\bsnm{Li}, \binits{W.}},
\bauthor{\bsnm{Cai}, \binits{X.}}:
\batitle{Empirical analysis of a scale-free railway network in china}.
\bjtitle{Physica A}
\bvolume{382}(\bissue{2}),
\bfpage{693}--\blpage{703}
(\byear{2007})
\end{barticle}
\endbibitem

\bibitem[\protect\citeauthoryear{Wang et~al.}{2020}]{Wang2020-dc}
\begin{barticle}
\bauthor{\bsnm{Wang}, \binits{L.-N.}},
\bauthor{\bsnm{Wang}, \binits{K.}},
\bauthor{\bsnm{Shen}, \binits{J.-L.}}:
\batitle{Weighted complex networks in urban public transportation: Modeling and testing}.
\bjtitle{Physica A}
\bvolume{545}(\bissue{123498}),
\bfpage{123498}
(\byear{2020})
\end{barticle}
\endbibitem

\bibitem[\protect\citeauthoryear{Liang et~al.}{2013}]{Liang2013-oy}
\begin{barticle}
\bauthor{\bsnm{Liang}, \binits{X.}},
\bauthor{\bsnm{Zhao}, \binits{J.}},
\bauthor{\bsnm{Dong}, \binits{L.}},
\bauthor{\bsnm{Xu}, \binits{K.}}:
\batitle{Unraveling the origin of exponential law in intra-urban human mobility}.
\bjtitle{Sci. Rep.}
\bvolume{3}(\bissue{1}),
\bfpage{2983}
(\byear{2013})
\end{barticle}
\endbibitem

\bibitem[\protect\citeauthoryear{Ji et~al.}{2017}]{Ji2017-vw}
\begin{barticle}
\bauthor{\bsnm{Ji}, \binits{Y.}},
\bauthor{\bsnm{Fan}, \binits{Y.}},
\bauthor{\bsnm{Ermagun}, \binits{A.}},
\bauthor{\bsnm{Cao}, \binits{X.}},
\bauthor{\bsnm{Wang}, \binits{W.}},
\bauthor{\bsnm{Das}, \binits{K.}}:
\batitle{Public bicycle as a feeder mode to rail transit in china: The role of gender, age, income, trip purpose, and bicycle theft experience}.
\bjtitle{Int. J. Sustain. Transp.}
\bvolume{11}(\bissue{4}),
\bfpage{308}--\blpage{317}
(\byear{2017})
\end{barticle}
\endbibitem

\bibitem[\protect\citeauthoryear{Zhang and Mi}{2018}]{Zhang2018-va}
\begin{barticle}
\bauthor{\bsnm{Zhang}, \binits{Y.}},
\bauthor{\bsnm{Mi}, \binits{Z.}}:
\batitle{Environmental benefits of bike sharing: A big data-based analysis}.
\bjtitle{Appl. Energy}
\bvolume{220},
\bfpage{296}--\blpage{301}
(\byear{2018})
\end{barticle}
\endbibitem

\bibitem[\protect\citeauthoryear{Oja et~al.}{2011}]{Oja2011-xy}
\begin{barticle}
\bauthor{\bsnm{Oja}, \binits{P.}},
\bauthor{\bsnm{Titze}, \binits{S.}},
\bauthor{\bsnm{Bauman}, \binits{A.}},
\bauthor{\bsnm{Geus}, \binits{B.}},
\bauthor{\bsnm{Krenn}, \binits{P.}},
\bauthor{\bsnm{Reger-Nash}, \binits{B.}},
\bauthor{\bsnm{Kohlberger}, \binits{T.}}:
\batitle{Health benefits of cycling: a systematic review: Cycling and health}.
\bjtitle{Scand. J. Med. Sci. Sports}
\bvolume{21}(\bissue{4}),
\bfpage{496}--\blpage{509}
(\byear{2011})
\end{barticle}
\endbibitem

\bibitem[\protect\citeauthoryear{DeMaio}{2009}]{DeMaio2009-ft}
\begin{barticle}
\bauthor{\bsnm{DeMaio}, \binits{P.}}:
\batitle{Bike-sharing: History, impacts, models of provision, and future}.
\bjtitle{J. Public Trans.}
\bvolume{12}(\bissue{4}),
\bfpage{41}--\blpage{56}
(\byear{2009})
\end{barticle}
\endbibitem

\bibitem[\protect\citeauthoryear{Pucher et~al.}{2011}]{Pucher2011-qt}
\begin{barticle}
\bauthor{\bsnm{Pucher}, \binits{J.}},
\bauthor{\bsnm{Buehler}, \binits{R.}},
\bauthor{\bsnm{Seinen}, \binits{M.}}:
\batitle{Bicycling renaissance in north america? an update and re-appraisal of cycling trends and policies}.
\bjtitle{Transp. Res. Part A Policy Pract.}
\bvolume{45}(\bissue{6}),
\bfpage{451}--\blpage{475}
(\byear{2011})
\end{barticle}
\endbibitem

\bibitem[\protect\citeauthoryear{Fishman et~al.}{2013}]{Fishman2013-eo}
\begin{barticle}
\bauthor{\bsnm{Fishman}, \binits{E.}},
\bauthor{\bsnm{Washington}, \binits{S.}},
\bauthor{\bsnm{Haworth}, \binits{N.}}:
\batitle{Bike share: A synthesis of the literature}.
\bjtitle{Transp. Rev.}
\bvolume{33}(\bissue{2}),
\bfpage{148}--\blpage{165}
(\byear{2013})
\end{barticle}
\endbibitem

\bibitem[\protect\citeauthoryear{Zhou}{2015}]{Zhou2015-ky}
\begin{barticle}
\bauthor{\bsnm{Zhou}, \binits{X.}}:
\batitle{Understanding spatiotemporal patterns of biking behavior by analyzing massive bike sharing data in chicago}.
\bjtitle{PLoS One}
\bvolume{10}(\bissue{10}),
\bfpage{0137922}
(\byear{2015})
\end{barticle}
\endbibitem

\bibitem[\protect\citeauthoryear{Kim}{2023}]{Kim2023-pd}
\begin{barticle}
\bauthor{\bsnm{Kim}, \binits{K.}}:
\batitle{Discovering spatiotemporal usage patterns of a bike-sharing system by type of pass: a case study from seoul}.
\bjtitle{Transportation (Amst.)}
\bvolume{51}(\bissue{4}),
\bfpage{1}--\blpage{35}
(\byear{2023})
\end{barticle}
\endbibitem

\bibitem[\protect\citeauthoryear{Basak et~al.}{2023}]{Basak2023-gf}
\begin{barticle}
\bauthor{\bsnm{Basak}, \binits{E.}},
\bauthor{\bsnm{Al~Balawi}, \binits{R.}},
\bauthor{\bsnm{Fatemi}, \binits{S.}},
\bauthor{\bsnm{Tafti}, \binits{A.}}:
\batitle{When crisis hits: Bike-sharing platforms amid the covid-19 pandemic}.
\bjtitle{PLoS One}
\bvolume{18}(\bissue{4}),
\bfpage{0283603}
(\byear{2023})
\end{barticle}
\endbibitem

\bibitem[\protect\citeauthoryear{}{}]{citibike_data}
\begin{botherref}
Citi Bike System Data.
\url{https://citibikenyc.com/system-data}.
Accessed: October 2023
\end{botherref}
\endbibitem

\bibitem[\protect\citeauthoryear{}{}]{santander_data}
\begin{botherref}
Santander Cycles Data.
\url{https://cycling.data.tfl.gov.uk/}.
Accessed: August 2024
\end{botherref}
\endbibitem

\bibitem[\protect\citeauthoryear{}{}]{docomo_data}
\begin{botherref}
Docomo Bike Share Data.
\url{https://ckan.odpt.org/dataset/c_bikeshare_gbfs-d-bikeshare}.
Accessed: October 2023
\end{botherref}
\endbibitem

\bibitem[\protect\citeauthoryear{}{}]{bluebikes_data}
\begin{botherref}
Bluebikes Data.
\url{https://s3.amazonaws.com/hubway-data/index.html}.
Accessed: August 2024
\end{botherref}
\endbibitem

\bibitem[\protect\citeauthoryear{}{}]{divvy_data}
\begin{botherref}
Divvy Data.
\url{https://www.divvybikes.com/system-data}.
Accessed: October 2023
\end{botherref}
\endbibitem

\bibitem[\protect\citeauthoryear{}{}]{capital_data}
\begin{botherref}
Capital BikeShare Data.
\url{https://capitalbikeshare.com/system-data}.
Accessed: October 2023
\end{botherref}
\endbibitem

\bibitem[\protect\citeauthoryear{Gebhart and Noland}{2014}]{Gebhart2014-tk}
\begin{barticle}
\bauthor{\bsnm{Gebhart}, \binits{K.}},
\bauthor{\bsnm{Noland}, \binits{R.B.}}:
\batitle{The impact of weather conditions on bikeshare trips in washington, {DC}}.
\bjtitle{Transportation (Amst.)}
\bvolume{41}(\bissue{6}),
\bfpage{1205}--\blpage{1225}
(\byear{2014})
\end{barticle}
\endbibitem

\bibitem[\protect\citeauthoryear{}{}]{meteostat}
\begin{botherref}
Meteostat.
\url{https://meteostat.net/en}.
Accessed: August 2024
\end{botherref}
\endbibitem

\bibitem[\protect\citeauthoryear{Loaiza-Monsalve and Riascos}{2019}]{Loaiza-Monsalve2019-hu}
\begin{barticle}
\bauthor{\bsnm{Loaiza-Monsalve}, \binits{D.}},
\bauthor{\bsnm{Riascos}, \binits{A.P.}}:
\batitle{Human mobility in bike-sharing systems: Structure of local and non-local dynamics}.
\bjtitle{PLoS One}
\bvolume{14}(\bissue{3}),
\bfpage{0213106}
(\byear{2019})
\end{barticle}
\endbibitem

\bibitem[\protect\citeauthoryear{Blondel et~al.}{2008}]{Blondel2008}
\begin{barticle}
\bauthor{\bsnm{Blondel}, \binits{V.D.}},
\bauthor{\bsnm{Guillaume}, \binits{J.-L.}},
\bauthor{\bsnm{Lambiotte}, \binits{R.}},
\bauthor{\bsnm{Lefebvre}, \binits{E.}}:
\batitle{Fast unfolding of communities in large networks}.
\bjtitle{Journal of statistical mechanics: theory and experiment}
\bvolume{2008}(\bissue{10}),
\bfpage{10008}
(\byear{2008})
\end{barticle}
\endbibitem

\end{itemize}


\end{document}